\documentclass[twocolumn]{revtex4}

\usepackage{graphicx}
\usepackage{dcolumn}
\usepackage{bm}
\usepackage{color}
\usepackage{amssymb,amsmath}
\newcommand{\mathsym}[1]{{}}

\input epsf
\begin{document}

\title{\Large Spherical Top-Hat Collapse of Viscous Modified Chaplygin Gas
in Einstein's Gravity and Loop Quantum Cosmology}

\author{\textbf{Ujjal Debnath}$^1$\footnote{ujjaldebnath@gmail.com, ujjal@iucaa.ernet.in}
and \textbf{Mubasher Jamil}$^2$\footnote{mjamil@sns.nust.edu.pk}}
\affiliation{$^1$Department of Mathematics, Indian Institute of
Engineering Science and Technology, Shibpur, Howrah-711103,
India.\\
$^2$Department of Mathematics, School of Natural Sciences
(SNS),National University of Sciences and Technology (NUST), H-12,
Islamabad, Pakistan.}

\begin{abstract}
{\bf Abstract:} In this work, we focus on the collapse of a
spherically symmetric perturbation, with a classical top-hat
profile, to study the nonlinear evolution of only viscous modified
Chaplygin gas (VMCG) perturbations in Einstein's gravity as well
as in loop quantum Cosmology (LQC). In the perturbed region, we
have investigated the natures of equation of state parameter,
square speed of sound and another perturbed quantities. The
results have been analyzed by numerical and graphical
investigations.
\end{abstract}

\maketitle

\section{Introduction}

Recent years have witnessed the emergence of the idea of
accelerating Universe and due to some observational results
\cite{Riess,Perlm} it is now established that the Universe is
accelerating. This acceleration is caused by some unknown matter
dubbed as dark energy. This dark energy has the positive energy
density and strong negative pressure that satisfies equation of
state parameter $w=p/\rho<-1/3$. The present acceleration is also
confirmed by other observations like large scale structure
\cite{Teg}, CMBR \cite{Sper1} and WMAP \cite{Briddle,Sper2}. The
observations predict that the present Universe occupies $\sim$ 4\%
ordinary matter, $\sim$ 74\% dark energy and $\sim$ 22\% dark
matter. The most simple candidate of dark energy is the cosmological
constant $\Lambda$ which satisfies the EoS parameter $w=-1$
\cite{Sahni}. Another candidates of dark energy are quintessence
(where EoS parameter satisfies $-1<w<-1/3$) \cite{Cald} and phantom
(where EoS parameter satisfies $w<-1$) \cite{Cald1}. A brief review
of dark energy models is found in the ref. \cite{Cop}. The unified
dark fluid (UDF) model \cite{Hu,Kamen} was investigated extensively
in the recent years. The main features of the UDF model are that it
combines cold dark matter and dark energy and that it behaves like
the cold dark matter and the dark energy at an early epoch and a
late time respectively however the approach remains purely
phenomenological. In recent years, an increasing number of
cosmological observations suggest that our universe is filled with
imperfect fluid which contains bulk viscosity which has negative
pressure \cite{Bala,Zim}. The viscous generalized Chaplygin gas
model is one of the suitable candidate of unified dark energy and
cold dark matter as a unique imperfect dark fluid \cite{Li0,Li00}.

While dark energy may be modeled as a fluid with negative pressure
acting against gravitational collapse, dark matter (in its cold
version) is a dust-like fluid with no pressure, therefore enhancing
the collapse of matter perturbations \cite{Feng,Garr}. It is very
important to investigate the evolutions of density perturbations
within realistic cosmological model (for recent reviews, see for
instance \cite{rev} and references therein). Fabris et al
\cite{fabris} studied the evolution of density perturbations in a
universe dominated by the Chaplygin gas. Their model shows the
required density contrast observed in large scale structure of the
universe, however their approach remained Newtonian. Similar
investigations were performed by Carturan and Finelli for the
generalized Chaplygin gas \cite{fin} (see also \cite{works}).
Fernandes et al \cite{Fern} have investigated the non-linear
collapse of generalized Chaplygin gas in the frame of spherical
top-hat collapse (STHC). Recently Carames \cite{car} investigated
the spherical collapse model using viscous generalized Chaplygin
gas. Li et al \cite{Li0,Li00} have extended the above work by
considering bulk viscosity in the general Chaplygin gas model.
Besides the parameter $\alpha$, they have also analyzed the effect
of bulk viscosity on the structure formation of the variable
generalized Chaplygin gas model which has a spherically symmetric
perturbation.

In our work, we focus on the collapse of a spherically symmetric
perturbation, with a classical top-hat profile, to study the
nonlinear evolution of viscous modified Chaplygin gas perturbations
in Einstein's gravity as well as in loop quantum Cosmology in
sections II and III respectively. The conclusion is present in the
last section IV.

\section{Spherical Top-hat collapse model of viscous modified Chaplygin gas in Einstein's gravity}

\subsection{Basic Equations}

We consider a flat Friedmann Robertson-Walker (FRW) universe
described by the following metric
\begin{equation}
ds^{2}=-dt^{2}+a^{2}(t)\left[dr^{2}+r^{2}d\theta^{2}+r^{2}\sin^{2}\theta
d\phi^{2} \right].
\end{equation}
Here $a(t)$ is the scale factor of the universe. We assume that
the spacetime is filled with only one component fluid having a
bulk viscosity. The Einstein's field equations are given by
\begin{equation}
H^{2}=\frac{8\pi G}{3}~\rho,
\end{equation}
and
\begin{equation}
\dot{H}=-4\pi G(\rho+p),
\end{equation}
where $H(=\frac{\dot{a}}{a})$ is the Hubble parameter, $\rho$ is
the energy density and the effective pressure $p$ can be expressed
as follows
\begin{equation}
p=p_{d}+\Pi,
\end{equation}
which is the sum of the equilibrium pressure $p_{de}$ and the bulk
pressure $\Pi=-\xi u^\gamma_{;\gamma}$, where $u^\gamma$ is the four
velocity of the fluid and $\xi$ is the coefficient of bulk viscosity
and is a function of energy density. The first attempts at creating
a viscosity theory of relativistic fluids were executed by Eckart
\cite{Eckart} and Landau and Lifshitz \cite{Landau} who considered
only a first-order deviation from equilibrium. The bulk viscous
pressure $\Pi$ is represented by the Eckart's expression which is
proportional to the Hubble parameter $H$ with proportionality factor
identified as the bulk viscosity coefficient $\xi$ which is defined
by $\xi=\xi_{0}\rho^{\nu}$, $\xi_{0}$ and $\nu$ are constants. For
simplicity, choosing $\nu=\frac{1}{2}$, $\Pi$ can be written as
\cite{Li0}
\begin{equation}
\Pi=-3\xi_{0}H\sqrt{\rho}.
\end{equation}

The Chaplygin gas is generally characterized by the equation of
state (pressure is inversely proportional to the energy density):
$p=-A/\rho$. This equation of state has an interesting connection
with the $d$-branes which are expressed via Nambu-Goto action
\cite{fabris}. It also enjoys connections with the Newtonian
hydrodynamical equations. Further the Eddington-Born-Infeld model
can be seen as an affine connection version for the Chaplygin gas
approach \cite{rod}. The Chaplygin gas has been extensively studied
within the unified dark energy-dark matter models as well
\cite{jamil}. Therefore we choose the equation of state for modified
Chaplygin gas as \cite{Deb}
\begin{equation}
p_{d}=A\rho-\frac{B}{\rho^{\alpha}},
\end{equation}
where $A$, $B$ and $\alpha$ are constants which are constrained by
the astrophysical data (for recent constraints, see \cite{paul}).
The modified Chaplygin gas satisfactorily accommodates an
accelerating phase followed by a matter dominated phase of the
universe. It is also consistent with the observational studies
dealing with the large scale structure \cite{paul}. Hence combining
the equations (4) to (6), we get
\begin{equation}
p=A\rho-\frac{B}{\rho^{\alpha}}-3\xi_{0}H\sqrt{\rho},
\end{equation}
which is called viscous modified Chaplygin gas (VMCG) \cite{jamil1}.
Hence by choosing the above equation of state, we extend previous
works dealing with viscous linear equation of state ($B=0$) and
viscous generalized Chaplygin gas ($A=0$) \cite{Li0,Li00}. It may
also be termed as a `viscous unified dark fluid'. The conservation
equation is given by
\begin{equation}
\dot{\rho}+3H(\rho+p)=0
\end{equation}
Now using equations (7) and (8), we obtain the solution
\begin{equation}
\rho=\left(\frac{B}{1+A-\sqrt{3}~\xi_{0}}+\frac{C}{a^{3(1+\alpha)(1+A-\sqrt{3}~\xi_{0})}}
\right)^{\frac{1}{1+\alpha}},
\end{equation}
where $C$ is a constant of integration. Now using the redshift
formula $z=\frac{1}{a}-1$ (choosing $a_{0}=1$), equation (9) can be
re-written in the form:
\begin{equation}
\rho(z)=\rho_{0}\left[A_{s}+(1-A_{s})(1+z)^{3(1+\alpha)(1+A-\sqrt{3}\xi_{0})}
\right]^{\frac{1}{1+\alpha}},
\end{equation}
where $\rho_{0}$ is the present value of the density and
$A_{s}=\frac{B}{(1+A-\sqrt{3}\xi_{0})C+B}$~with $0< A_{s}< 1$ and
$1+A>\sqrt{3}\xi_{0}$. Hence the Hubble parameter is obtained as
\begin{eqnarray*}
H(z)=H_{0}\left[\Omega_{0}\left\{A_{s}+(1-A_{s})\times
\right.\right.
\end{eqnarray*}
\begin{eqnarray}
\left.\left. (1+z)^{3(1+\alpha)(1+A-\sqrt{3}\xi_{0})}
\right\}^{\frac{1}{1+\alpha}}\right]^{\frac{1}{2}}
\end{eqnarray}
where $\Omega_{0}=\frac{8\pi G\rho_{0}}{3H_{0}^{2}}$ is the
present value of dimension density parameter and $H_{0}\sim
72km~s^{-1}Mpc^{-1}$. The equation of state parameter is given by
\begin{equation}
w=\frac{p}{\rho}=A-\frac{B}{\rho^{1+\alpha}}-\frac{3\xi_{0}H}{\sqrt{\rho}},
\end{equation}
while the adiabatic sound speed reads
\begin{equation}
c_{s}^{2}=\frac{\partial p}{\partial\rho}=A+\frac{\alpha
B}{\rho^{1+\alpha}}-\frac{\xi_{0}H}{2\sqrt{\rho}}.
\end{equation}
Note that the Chaplygin gas model is a dynamical dark energy model
as well and dark energy perturbations play crucial role in dynamical
dark energy models as well \cite{chan}.

\subsection{Spherical top-hat collapse model}

The spherical collapse (SC) which provides a way to glimpse into
the nonlinear regime of perturbation theory was introduced firstly
by Gunn and Gott \cite{Gunn}. The SC describes the evolution of a
spherically symmetric perturbation embedded in a homogeneous
background, which can be static, expanding or collapsing. One
assumes a spherical `top-hat' profile for the perturbed region,
i.e., a spherically symmetric perturbation in some region of space
with constant density. Following the assumption of a top-hat
profile, namely the density perturbation is uniform throughout the
collapse, so the evolution of perturbation is only time-dependent.

The perturbed quantities $\rho_{c}$ and $p_{c}$ are related to their
background counterparts by $\rho_{c}=\rho+\delta\rho$ and
$p_{c}=p+\delta p$. Now the perturbed equation of state $w_{c}$ is
given by \cite{Fern,Li0,Li00}
\begin{eqnarray*}
w_{c}=\frac{p+\delta
p}{\rho+\delta\rho}=\frac{w+c_{e}^{2}\delta}{1+\delta}~~~~~~~~~~~~~~~
\end{eqnarray*}
\begin{equation}
=\frac{1}{1+\delta}
\left(A-\frac{B}{\rho^{1+\alpha}}-\frac{3\xi_{0}H}{\sqrt{\rho}}\right)
+\frac{c_{e}^{2}\delta}{1+\delta}
\end{equation}
where $\delta=\frac{\delta\rho}{\rho}$ and perturbed square of the
sound speed is given by
\begin{equation}
c_{e}^{2}=\frac{\delta
p}{\delta\rho}=\frac{p_{c}-p}{\rho_{c}-\rho}=A+\frac{B[(1+\delta)^{\alpha}-1
]}{\delta(\delta+1)^{\alpha}\rho^{1+\alpha}
}-\frac{3\xi_{0}H}{(\sqrt{1+\delta}+1)\sqrt{\rho}}.
\end{equation}
In the spherical top-hat collapse model, the background evolution
equations are still in the forms \cite{Fern,Li0,Li00}:
\begin{equation}
\dot{\rho}=-3H(\rho+p),
\end{equation}
\begin{equation}
\frac{\ddot{a}}{a}=-\frac{4\pi G}{3}(\rho+3p)
\end{equation}

For the perturbed region, the basic equations which depend on
local quantities can be written as
\begin{equation}
\dot{\rho}_{c}=-3h(\rho_{c}+p_{c}),
\end{equation}
\begin{equation}
\frac{\ddot{r}}{r}=-\frac{4\pi G}{3}(\rho_{c}+3p_{c})
\end{equation}
where $h=\frac{\dot{r}}{r}$ is the local expansion rate and $r$ is
the local scale factor and furthermore, $h$ relates to local
expansion rate in the STHC framework \cite{Fern,Ab},
\begin{equation}
h=H+\frac{\theta}{3a}
\end{equation}
where $\theta\equiv\nabla.\vec{v}$ is the divergence of the
peculiar velocity $\vec{v}$.

After some calculations \cite{Fern,ma}, the dynamical equations of
density contrast $\delta$ and $\theta$ can be obtained in the
following forms:
\begin{equation}
\delta'=-\frac{3}{a}(c_{e}^{2}-w)\delta-[1+w+(1+c_{e}^{2})\delta]\frac{\theta}{a^{2}H}~,
\end{equation}
\begin{equation}
\theta'=-\frac{\theta}{a}-\frac{\theta^{2}}{3a^{2}H}-\frac{3H}{2}\Omega\delta(1+3c_{e}^{2})
\end{equation}
where, $\Omega=\frac{8\pi G\rho}{3H^{2}}$~ and $'$ represents the
derivative w.r.t. scale factor $a$. The above two equations can be
re-written in the forms:
\begin{equation}
\frac{d\delta}{dz}=\frac{3(c_{e}^{2}-w)\delta}{1+z}+[1+w+(1+c_{e}^{2})\delta]\frac{\theta}{H(z)}~,
\end{equation}
\begin{equation}
\frac{d\theta}{dz}=\frac{\theta}{1+z}+\frac{\theta^{2}}{3H(z)}+
\frac{3H(z)}{2(1+z)^{2}}\Omega\delta(1+3c_{e}^{2}).
\end{equation}

We have drawn the time varying parameters
$\delta,\theta,w,w_{c},c_{s}^{2},c_{e}^{2},h$ vs redshift $z$ in
figures 1-7 respectively for Einstein's gravity. In all the
figures we have taken the values of the parameters
$A=0.9,B=0.6,C=0.2,\xi_{0}=0.2,\alpha=0.5,H_{0}=72,\Omega_{0}=1$.
From fig.1 we observe that the parameter $\delta$ increases from 0
to the value $\sim 1.8$ as the redshift $z$ decreases from 5 to
$-1$. Similar nature happens for the quantity $\theta$ (increases
from 0 to $\sim 3.3$) which is shown in fig.2. The equation of
state parameter $w$ vs $z$ is drawn in figure 3. We see that $w$
decreases from 0.15 to $-1$. So the VMCG generates initially
normal fluid and after certain stage, it generates quintessence
dark energy but phantom barrier does not happen, but meanwhile the
perturbed equation of state parameter $w_{c}$ does not so change
and its value $\sim 0.9$ all over the time. The square speed of
sound $c_{s}^{2}$ for our VMCG fluid system first increases from
0.61056 for $z\sim 2$ and ultimately increases to the value
0.61059 at $z\sim -0.5$ (fig.5). But, perturbed square speed of
sound $c_{e}^{2}$ actually increases from 0.74 (at $z\sim 5$) to
0.9 after $z\sim 4.8$ (fig.6). Also the expansion rate $h$ vs $z$
has been drawn in figure 7. We observe that $h$ first increases
from 0 (from $z\sim 5$) to $\sim 60.1$ around $z\sim 0.4$ and
after that suddenly decreases to zero (nearly $z=-1$).

\begin{figure}
\includegraphics[height=2.1in]{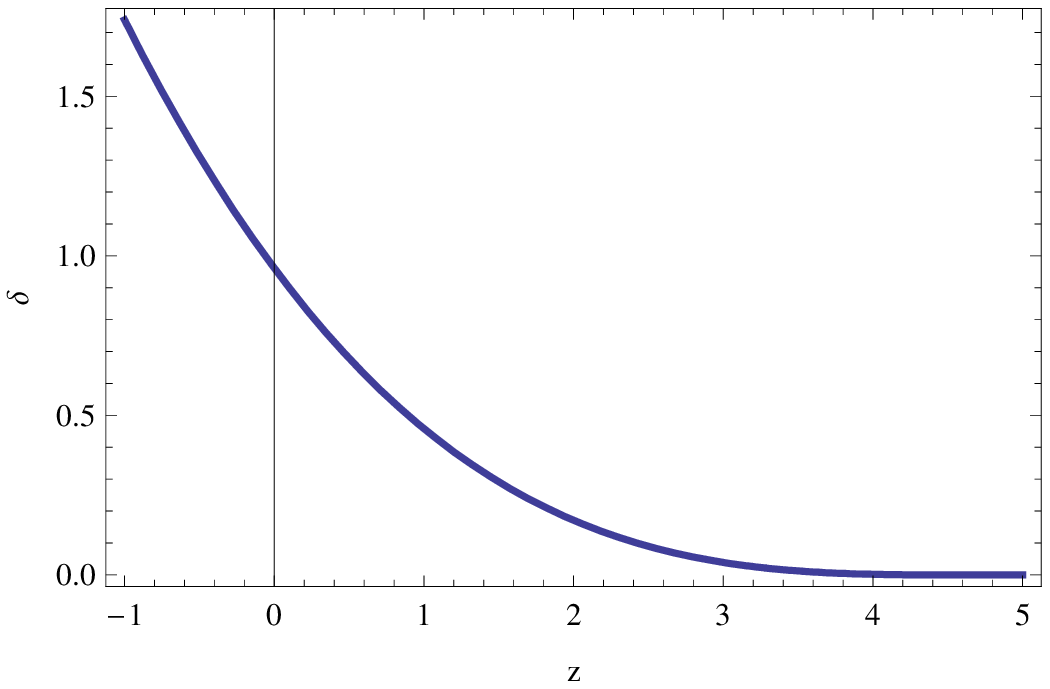}\\
~~~~~~~~~~~~~~~~~~~~~~~~~~~Fig.1~~~~~~~~~~~~~~~~~~\\
\vspace{4mm}
\includegraphics[height=2.1in]{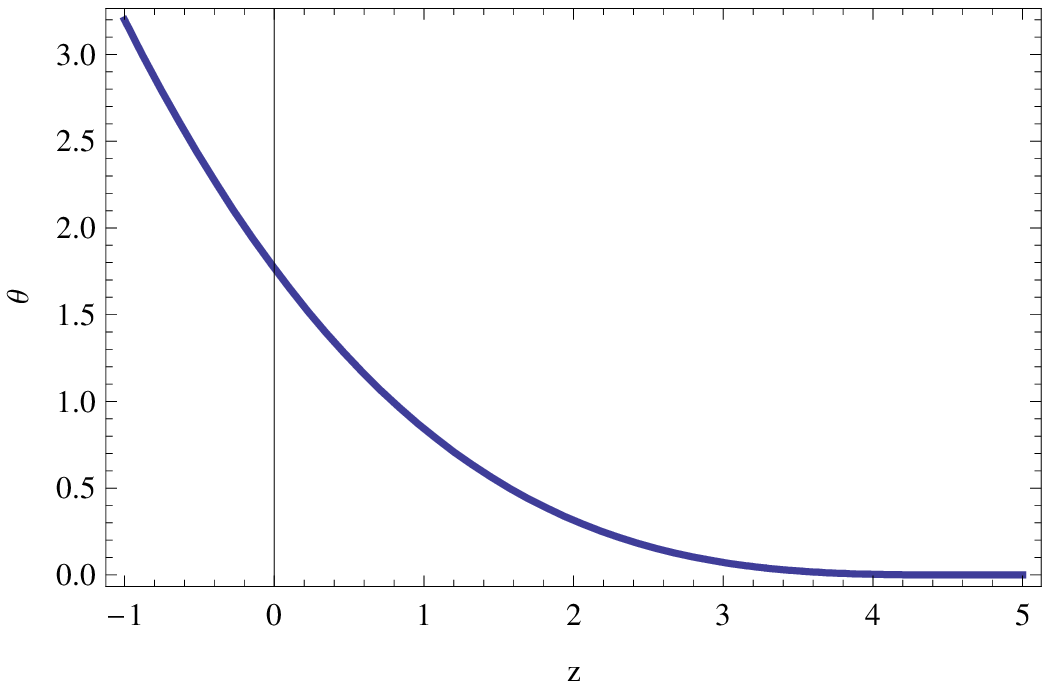}\\
~~~~~~~~~~~~~~~~~~~~~~~~~~~Fig.2~~~~~~~~~~~~~~~~~~\\
\vspace{4mm}
\includegraphics[height=2.1in]{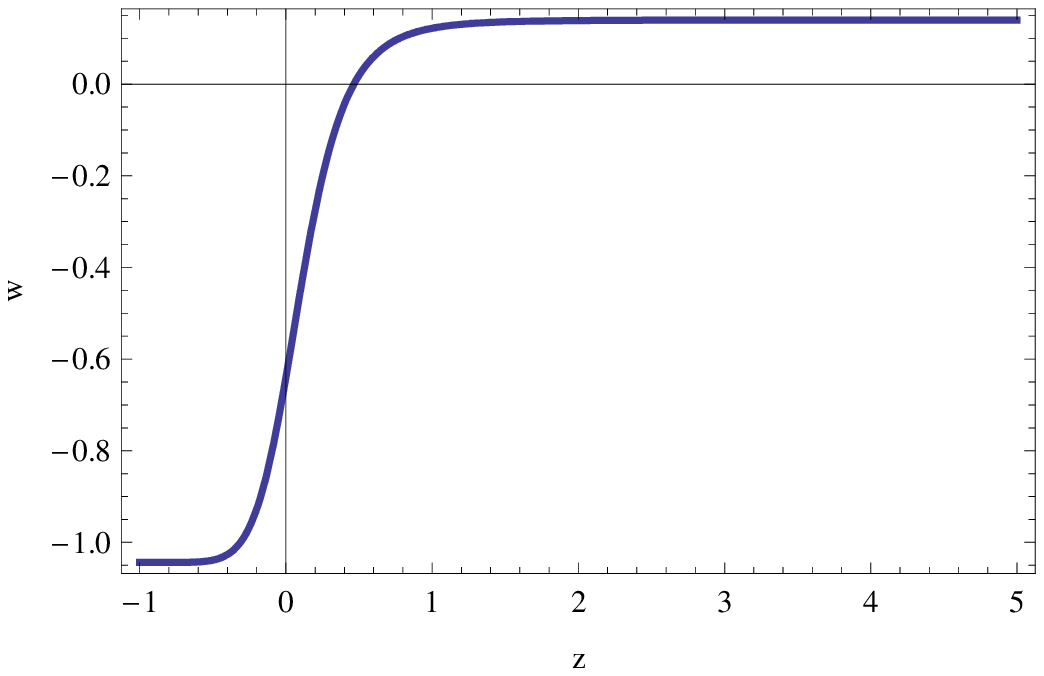}\\
~~~~~~~~~~~~~~~~~~~~~~~~~~~Fig.3~~~~~~~~~~~~~~~~~~\\
\vspace{4mm}
\includegraphics[height=2.1in]{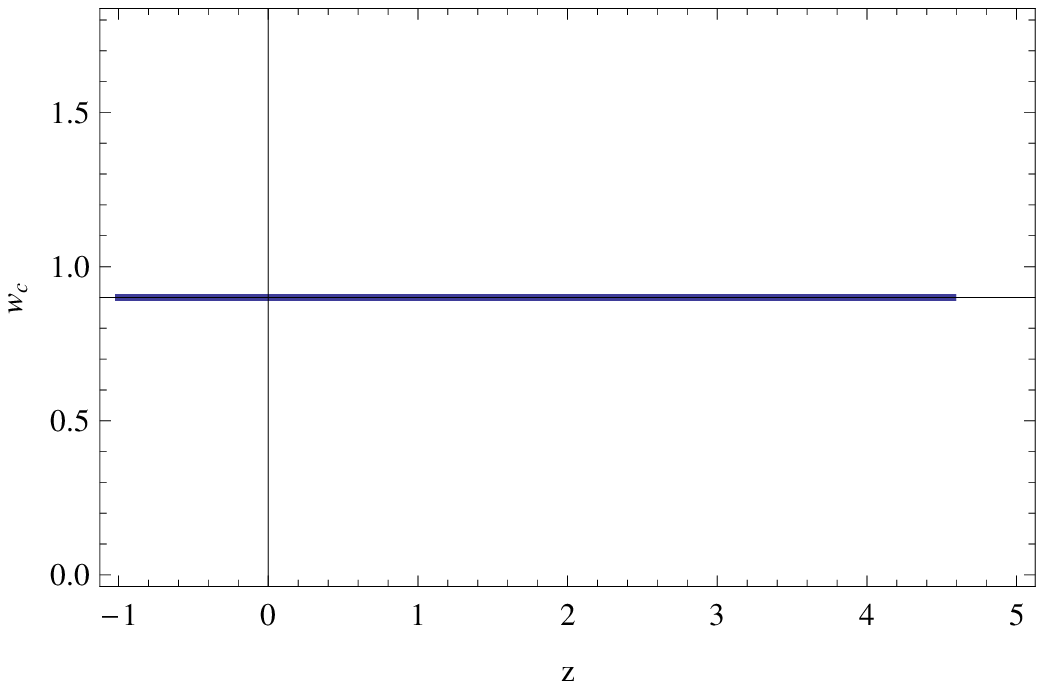}\\
~~~~~~~~~~~~~~~~~~~~~~~~~~~Fig.4~~~~~~~~~~~~~~~~~~\\
\vspace{4mm} Figs. 1-4: Plots of $\delta,\theta,w,w_{c}$ vs
redshift $z$ in Einstein's gravity. \vspace{.2in}
\end{figure}
\begin{figure}
\includegraphics[height=2.1in]{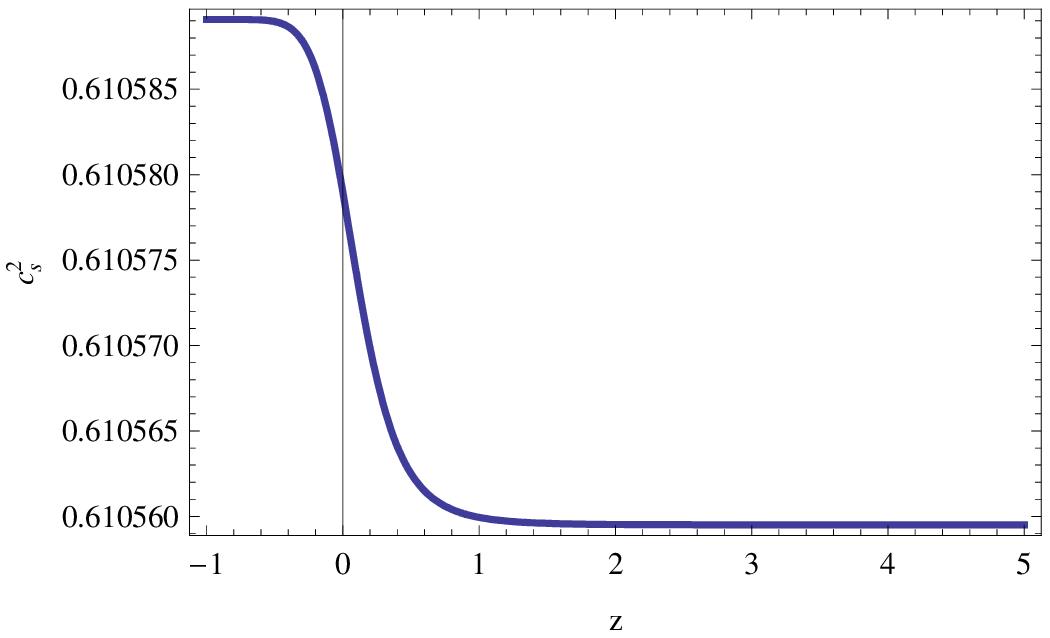}\\
~~~~~~~~~~~~~~~~~~~~~~~~~~~Fig.5~~~~~~~~~~~~~~~~~~\\
\vspace{4mm}
\includegraphics[height=2.1in]{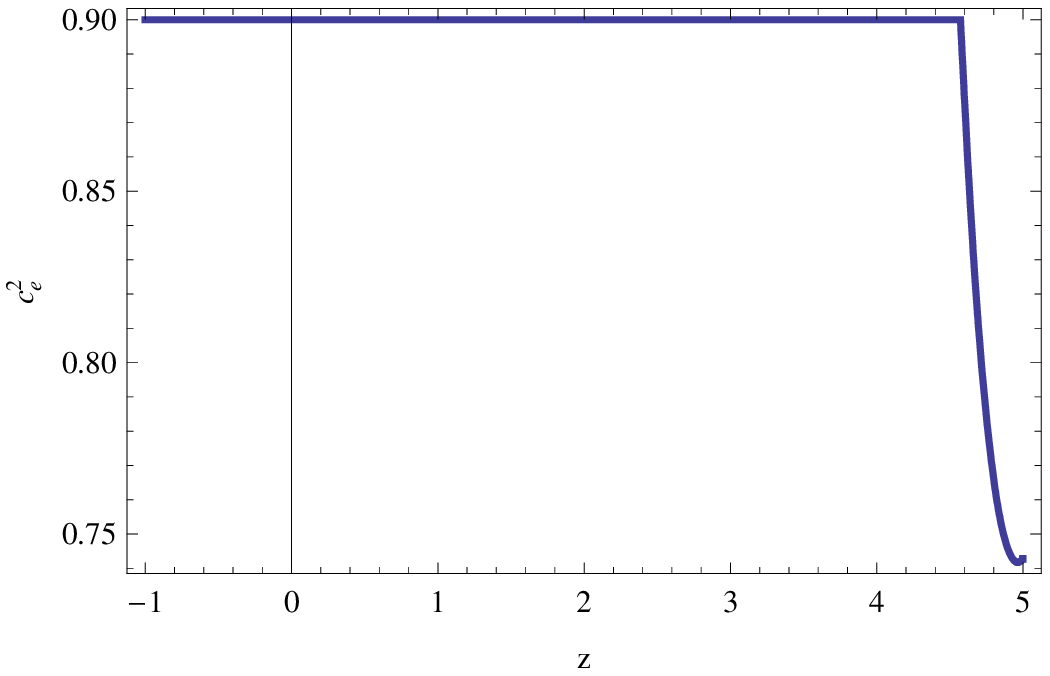}\\
~~~~~~~~~~~~~~~~~~~~~~~~~~~Fig.6~~~~~~~~~~~~~~~~~~\\
\vspace{4mm}
\includegraphics[height=2.1in]{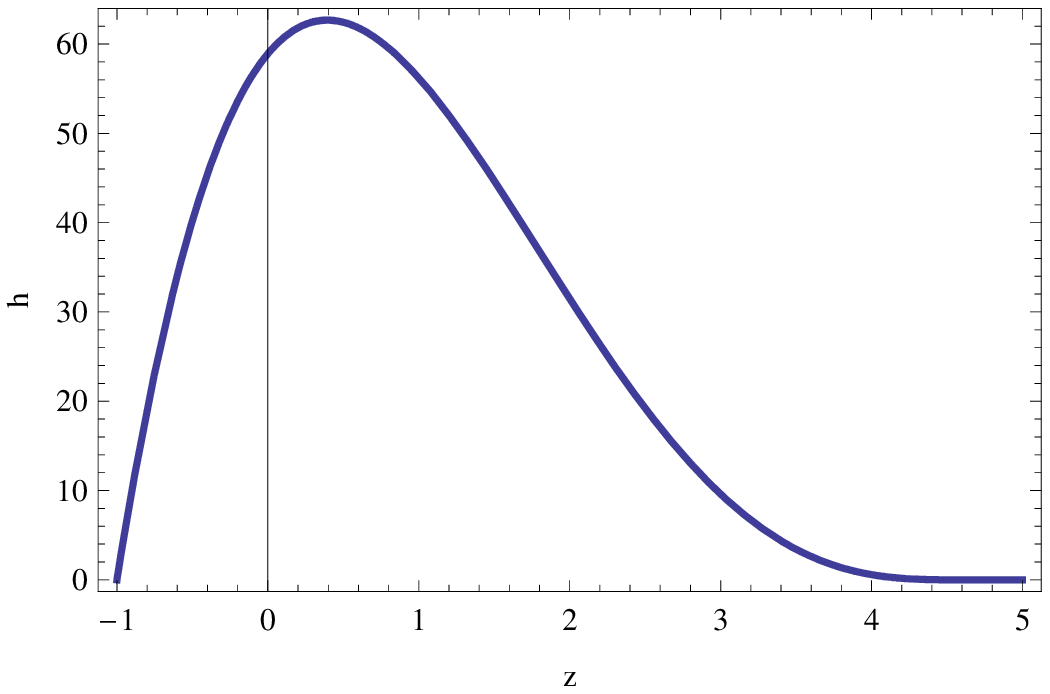}\\
~~~~~~~~~~~~~~~~~~~~~~~~~~~Fig.7~~~~~~~~~~~~~~~~~~\\
\vspace{4mm}Figs. 5-7: Plots of $c_{s}^{2},c_{e}^{2},h$ vs
redshift $z$ in Einstein's gravity. \vspace{.2in}
\end{figure}

\section{Spherical top-hat collapse model of viscous modified Chaplygin gas in Loop Quantum Cosmology}

Loop Quantum Gravity (LQG) is a canonical quantization of gravity
based upon Ashtekar connection variables. LQG is an important
frontier to explore the quantum gravity effects in cosmology. Some
of its implications includes the prediction of cosmic inflation in
the early universe \cite{find}, late time cosmic acceleration
\cite{vac} and primordial gravitational waves \cite{wv}. The field
equations of LQC admit attractor solutions which are of enormous
cosmological interest. The cosmological perturbation theory within
LQC has also been explored in \cite{per}. We here consider the flat
homogeneous and isotropic universe described by FRW metric, so the
modified Einstein's field equations in LQC are given by
\cite{Wu0,Chen0,Fu0}
\begin{equation}
H^{2}=\frac{8\pi G\rho}{3}\left(1-\frac{\rho}{\rho_{1}}\right),
\end{equation}
and
\begin{equation}
\dot{H}=-4\pi G(\rho+p)\left(1-\frac{2\rho}{\rho_{1}}\right),
\end{equation}
where $\rho_{1}=\frac{\sqrt{3}}{16\pi^{2}\gamma^{3}G^{2}\hbar}$ is
called the critical loop quantum density, $\gamma$ is the
dimensionless Barbero-Immirzi parameter. Now the fluid is
considered as viscous modified Chaplygin gas and in this case, the
density is given in equation (10) and in LQC, the Hubble parameter
is obtained as
\begin{eqnarray*}
H(z)=H_{0}\left[\Omega_{0}\left\{A_{s}+(1-A_{s})(1+z)^{3(1+\alpha)(1+A-\sqrt{3}\xi_{0})}
\right\}^{\frac{1}{1+\alpha}}\right]^{\frac{1}{2}}
\end{eqnarray*}
\begin{equation}
\times~   \left[1-\frac{3H_{0}^{2}\Omega_{0}}{\rho_{1}}
\left\{A_{s}+(1-A_{s})(1+z)^{3(1+\alpha)(1+A-\sqrt{3}\xi_{0})}
\right\}^{\frac{1}{1+\alpha}} \right]^{\frac{1}{2}}
\end{equation}
Similar to the above section, the equation (23) will be the same
and equation (24) modifies to the form
\begin{eqnarray*}
\frac{d\theta}{dz}=\frac{\theta}{1+z}+\frac{\theta^{2}}{3H(z)}+
\frac{3H(z)}{2(1+z)^{2}}\Omega\delta\times~~~~~~~~~~~~~
\end{eqnarray*}
\begin{equation}
\left[(1+3c_{e}^{2}) -\frac{6\Omega H^{2}(z)}{8\pi
G\rho_{1}}~\left\{3(1+\delta)(1+3c_{e}^{2}) +(3w-\delta+1)\right\}
\right]
\end{equation}
In this case also, the equation of state $w_{c}$ and square of the
sound speed are given in (14) and (15). We have drawn the time
varying parameters $\delta,\theta,w,w_{c},c_{s}^{2},c_{e}^{2},h$
vs redshift $z$ in figures 8-14 respectively for LQC. In all the
figures we have taken the values of the parameters
$A=0.9,B=0.6,C=0.2,\xi_{0}=0.2,\alpha=0.5,H_{0}=72,\Omega_{0}=1$.
From fig.8 we observe that the parameter $\delta$ decreases from a
certain value (around $z\sim 4$). Also the quantity $\theta$
(increases from 0 to $\sim 4.4$) during $z=5$ to $z=4.8$ and after
that $\theta$ decreases to zero (upto $z\sim -1$) which is shown
in fig.9. The equation of state parameter $w$ vs $z$ is drawn in
figure 10 which shows that $w$ decreases from some negative value
to $-0.84$ as $z$ decreases. So the VMCG in our considered LQC
model generates like quintessence dark energy but phantom barrier
does not happen. In the meanwhile the perturbed equation of state
parameter $w_{c}$ changes from some positive value $(<1)$ to $\sim
-0.9$ (fig.11). The square speed of sound $c_{s}^{2}$ for our VMCG
fluid system first decreases to the value 0.61 at $z\sim -0.5$
(fig.12). But, perturbed square speed of sound $c_{e}^{2}$
actually increases from some fractional value $(<0.5)$ to $\sim
0.02$ (fig.13). Also the expansion rate $h$ vs $z$ has been drawn
in figure 14. We observe that $h$ first increases from some
positive value (from $z\sim 5$) to $\sim 4$ around $z\sim 4.8$ and
after that it decreases to zero (nearly $z=-1$).

\begin{figure}
\includegraphics[height=2.1in]{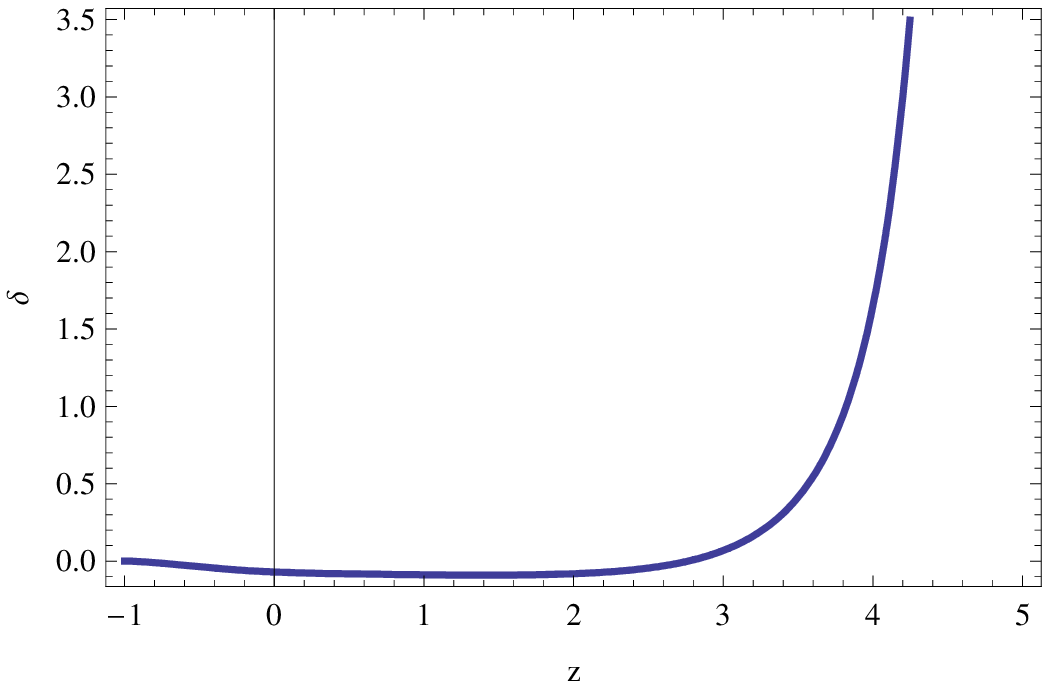}\\
~~~~~~~~~~~~~~~~~~~~~~~~~~~Fig.8~~~~~~~~~~~~~~~~~~\\
\vspace{4mm}
\includegraphics[height=2.1in]{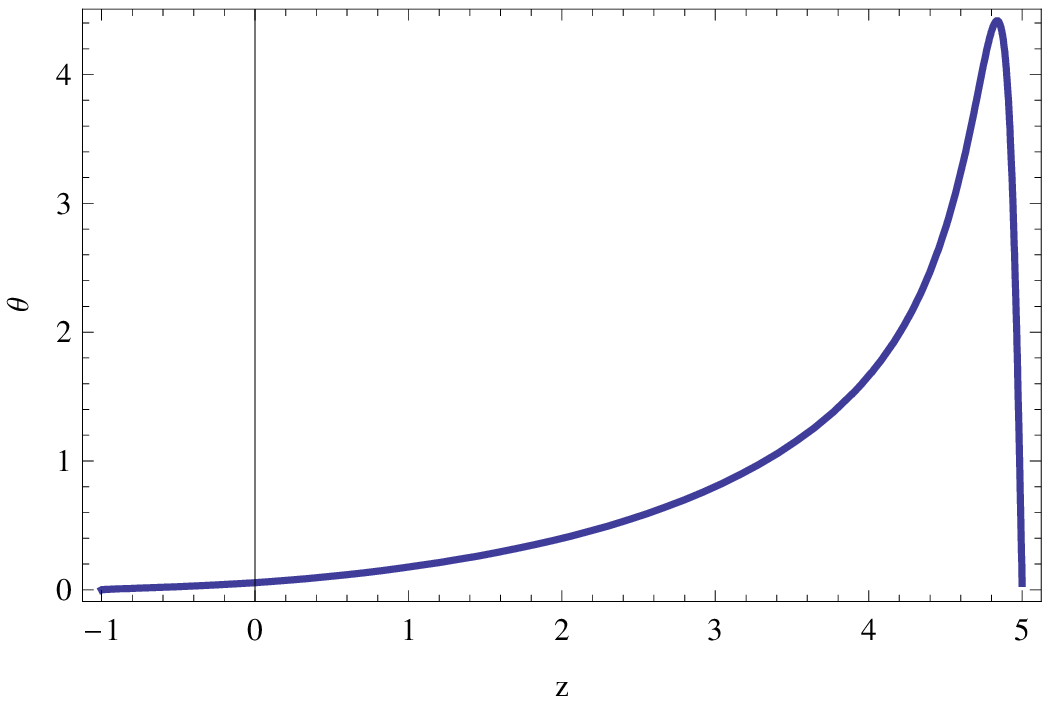}\\
~~~~~~~~~~~~~~~~~~~~~~~~~~~Fig.9~~~~~~~~~~~~~~~~~~\\
\vspace{4mm}
\includegraphics[height=2.1in]{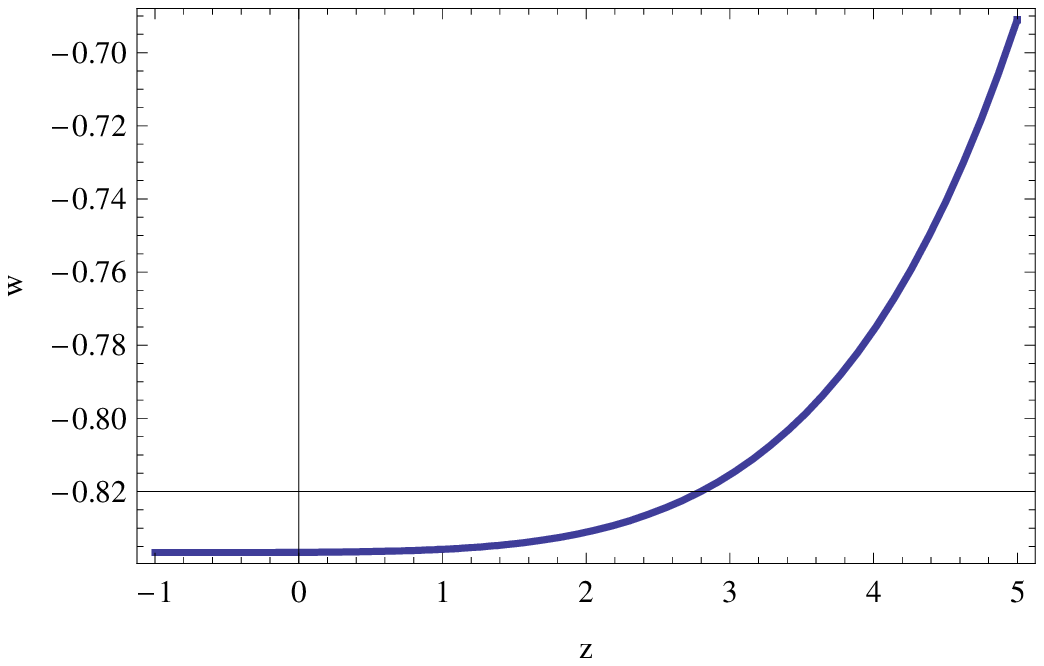}\\
~~~~~~~~~~~~~~~~~~~~~~~~~~~Fig.10~~~~~~~~~~~~~~~~~~\\
\vspace{4mm}
\includegraphics[height=2.1in]{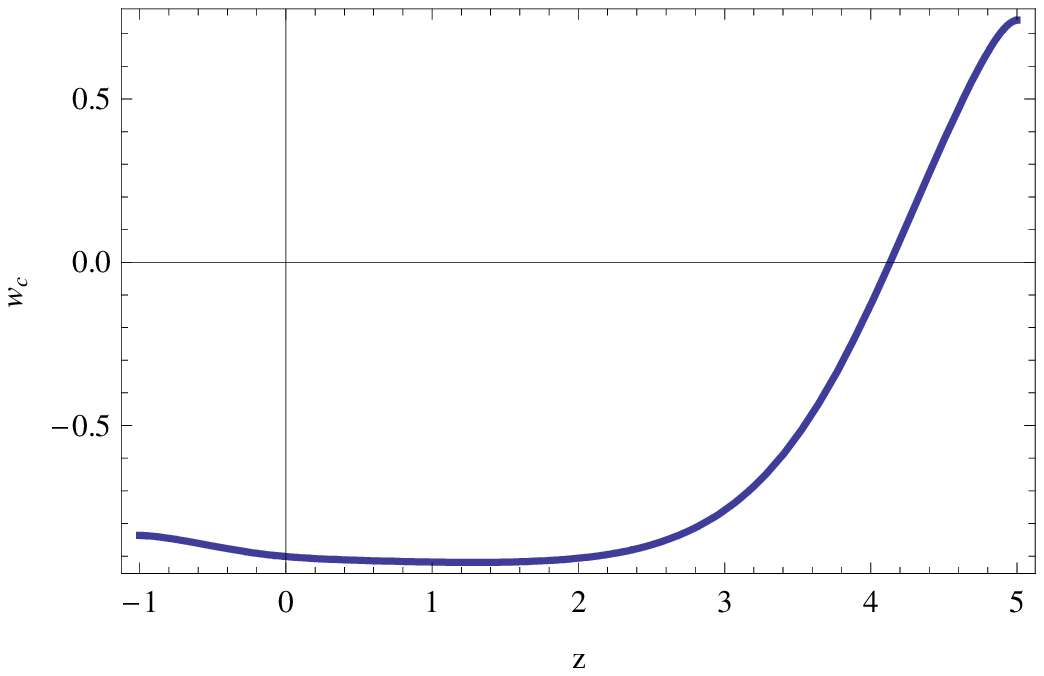}\\
~~~~~~~~~~~~~~~~~~~~~~~~~~~Fig.11~~~~~~~~~~~~~~~~~~\\
\vspace{4mm} Figs. 8-11: Plots of $\delta,\theta,w,w_{c}$ vs
redshift $z$ in LQC. \vspace{.2in}
\end{figure}
\begin{figure}
\includegraphics[height=2.1in]{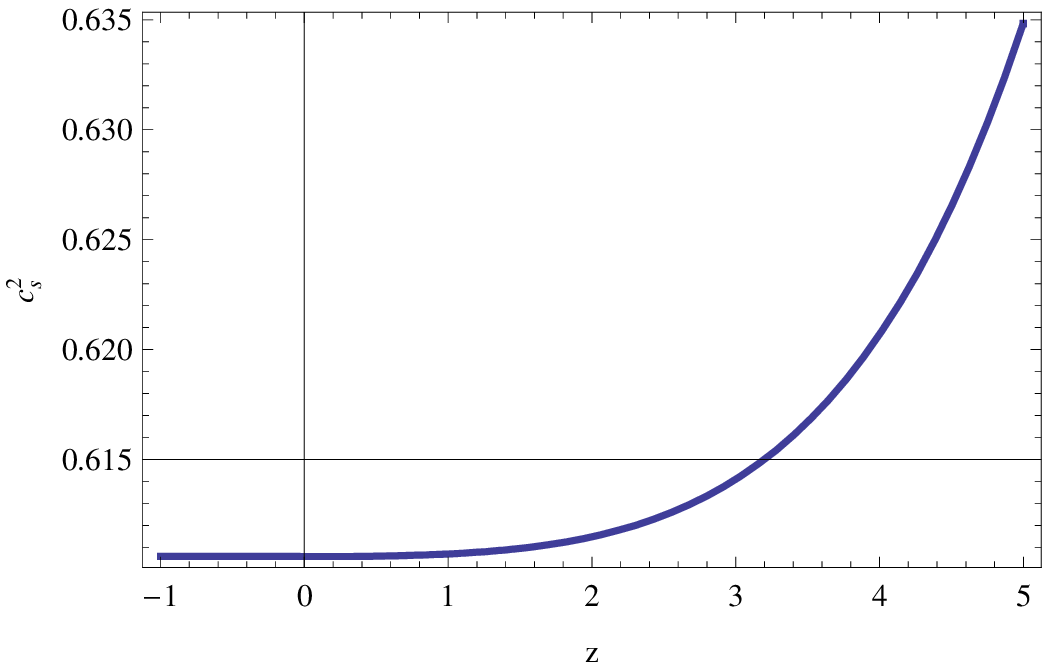}\\
~~~~~~~~~~~~~~~~~~~~~~~~~~~Fig.12~~~~~~~~~~~~~~~~~~\\
\vspace{4mm}
\includegraphics[height=2.1in]{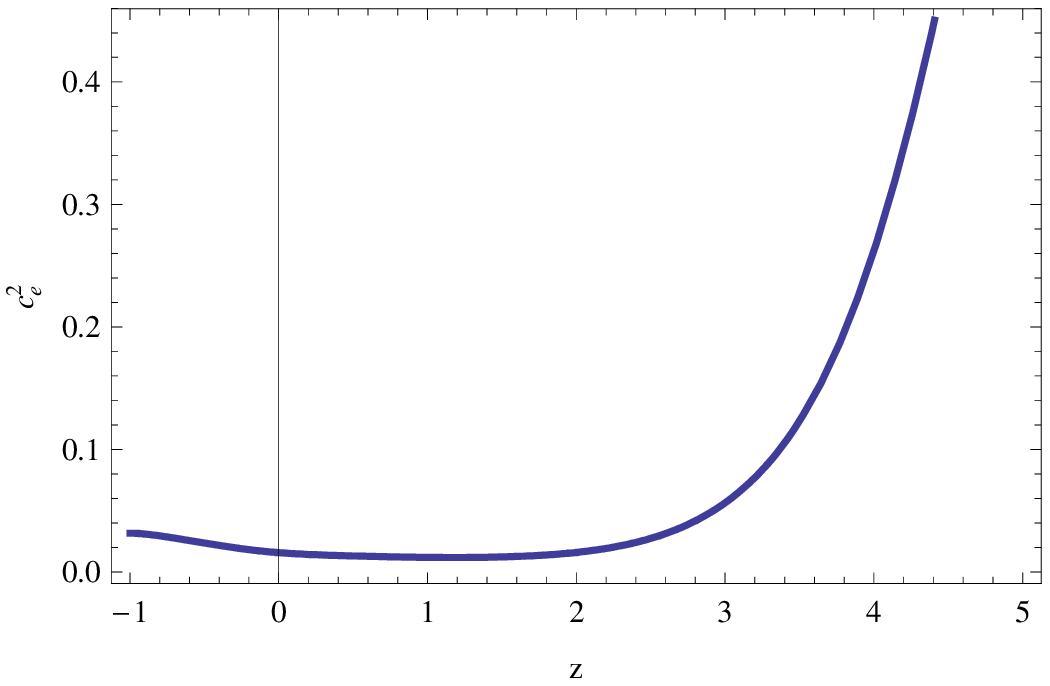}\\
~~~~~~~~~~~~~~~~~~~~~~~~~~~Fig.13~~~~~~~~~~~~~~~~~~\\
\vspace{4mm}
\includegraphics[height=2.1in]{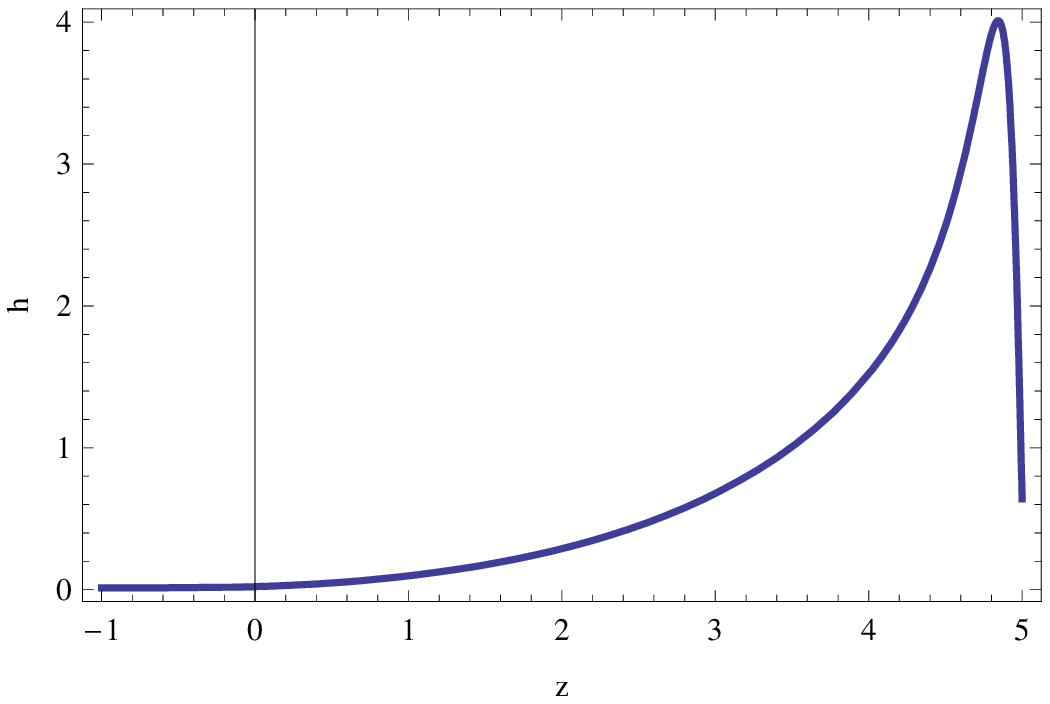}\\
~~~~~~~~~~~~~~~~~~~~~~~~~~~Fig.14~~~~~~~~~~~~~~~~~~\\
\vspace{4mm}Figs. 12-14: Plots of $c_{s}^{2},c_{e}^{2},h$ vs
redshift $z$ in LQC. \vspace{.2in}
\end{figure}

\section{Conclusions}

In this work, we mainly focused on the collapse of a spherically
symmetric perturbation, with a classical top-hat profile, to study
the nonlinear evolution of only viscous modified Chaplygin gas
(VMCG) perturbations in Einstein's gravity as well as in loop
quantum Cosmology (LQC). The background model is considered as
flat FRW metric. Since we know that modified Chaplygin gas (MCG)
is the unified model of dark matter and dark energy. So we have
not assumed any dark matter separately. In this occasion, we have
assumed viscous modified Chaplygin gas (VMCG) (which is also the
unified model as established) by including the viscosity term in
the equation of state in MCG. The density and the Hubble parameter
have been calculated in terms of redshift $z$ in both gravity
models. The equation of state parameter, square speed of sound
have also been calculated for our considered VMCG model in
Einstein's gravity and LQC also. Next step, the density
perturbation for our top-hat collapsing profile has been
investigated. In the perturbed region, we have investigated the
natures of equation of state parameter, square speed of sound and
another perturbed quantities like $\delta,~\theta,~h$ etc. The
dynamical equations of density contrast $\delta$ and $\theta$ have
been found in both gravity models. Analytically, it is not
possible to find out the natures of the perturbed quantities. So
numerically and graphically we analyzed the natures of the
perturbed quantities which are given as follows:\\

$\bullet{}$ $Einstein's ~Gravity$: We have drawn the time varying
parameters $\delta,\theta,w,w_{c},c_{s}^{2},c_{e}^{2},h$ vs
redshift $z$ in figures 1-7 respectively for Einstein's gravity.
In all the figures we have taken the values of the parameters
$A=0.9,B=0.6,C=0.2,\xi_{0}=0.2,\alpha=0.5,H_{0}=72,\Omega_{0}=1$.
From fig.1 we observe that the parameter $\delta$ increases from 0
to the value $\sim 1.8$ as the redshift $z$ decreases from 5 to
$-1$. Similar nature happens for the quantity $\theta$ (increases
from 0 to $\sim 3.3$) which is shown in fig.2. The equation of
state parameter $w$ vs $z$ is drawn in figure 3. We see that $w$
decreases from 0.15 to $-1$. So the VMCG generates initially
normal fluid and after certain stage, it generates quintessence
dark energy but phantom barrier does not happen, but meanwhile the
perturbed equation of state parameter $w_{c}$ does not so change
and its value $\sim 0.9$ all over the time. The square speed of
sound $c_{s}^{2}$ for our VMCG fluid system first increases from
0.61056 for $z\sim 2$ and ultimately increases to the value
0.61059 at $z\sim -0.5$ (fig.5). But, perturbed square speed of
sound $c_{e}^{2}$ actually increases from 0.74 (at $z\sim 5$) to
0.9 after $z\sim 4.8$ (fig.6). Also the expansion rate $h$ vs $z$
has been drawn in figure 7. We observe that $h$ first increases
from 0 (from $z\sim 5$) to $\sim 60.1$ around $z\sim 0.4$ and
after that suddenly decreases to zero (nearly $z=-1$).\\

$\bullet{}$ $LQC$: We have drawn the time varying parameters
$\delta,\theta,w,w_{c},c_{s}^{2},c_{e}^{2},h$ vs redshift $z$ in
figures 8-14 respectively for LQC. In all the figures we have
taken the values of the parameters
$A=0.9,B=0.6,C=0.2,\xi_{0}=0.2,\alpha=0.5,H_{0}=72,\Omega_{0}=1$.
From fig.8 we observe that the parameter $\delta$ decreases from a
certain value (around $z\sim 4$). Also the quantity $\theta$
(increases from 0 to $\sim 4.4$) during $z=5$ to $z=4.8$ and after
that $\theta$ decreases to zero (upto $z\sim -1$) which is shown
in fig.9. The equation of state parameter $w$ vs $z$ is drawn in
figure 10 which shows that $w$ decreases from some negative value
to $-0.84$ as $z$ decreases. So the VMCG in our considered LQC
model generates like quintessence dark energy but phantom barrier
does not happen. In the meanwhile the perturbed equation of state
parameter $w_{c}$ changes from some positive value $(<1)$ to $\sim
-0.9$ (fig.11). The square speed of sound $c_{s}^{2}$ for our VMCG
fluid system first decreases to the value 0.61 at $z\sim -0.5$
(fig.12). But, perturbed square speed of sound $c_{e}^{2}$
actually increases from some fractional value $(<0.5)$ to $\sim
0.02$ (fig.13). Also the expansion rate $h$ vs $z$ has been drawn
in figure 14. We observe that $h$ first increases from some
positive value (from $z\sim 5$) to $\sim 4$ around $z\sim 4.8$ and
after that it decreases to zero (nearly $z=-1$).\\

{\bf Acknowledgement:}\\

One of the author (UD) is thankful to IUCAA, Pune,
India for warm hospitality where part of the work was carried out.\\

\end{document}